\documentclass[letterpaper, 12pt]{article}[2000/05/19]
\usepackage[english]{babel}
\usepackage{amsfonts,amsmath,amssymb,amsthm,latexsym,amscd,mathrsfs}
\usepackage{ifthen,cite}
\usepackage[bookmarksnumbered=true]{hyperref}

\hypersetup{pdfpagetransition={Split}}

\newcommand{\evenhead}{Author \ name}
\newcommand{\oddhead}{Article \ name}
\newcommand{\theArticleName}{Article \ name}

\newcommand{\FirstPageHeading}[1]{\thispagestyle{empty}%
\noindent\raisebox{0pt}[0pt][0pt]{\makebox[\textwidth]{\protect\footnotesize \sf }}\par}

\newcommand{\ArticleName}[1]{\renewcommand{\theArticleName}{#1}\vspace{-2mm}\par\noindent {\LARGE\bf  #1\par}}
\newcommand{\Author}[1]{\vspace{5mm}\par\noindent {\Large  #1\par} \par\vspace{2mm}\par}
\newcommand{\Address}[1]{\vspace{2mm}\par\noindent {\it #1} \par}
\newcommand{\Email}[1]{\ifthenelse{\equal{#1}{}}{}{\par\noindent {\rm E-mail: }{\it  #1} \par}}
\newcommand{\URLaddress}[1]{\ifthenelse{\equal{#1}{}}{}{\par\noindent {\rm URL: }{\tt  #1} \par}}
\newcommand{\EmailD}[1]{\ifthenelse{\equal{#1}{}}{}{\par\noindent {$\phantom{\dag}$~\rm E-mail: }{\it  #1} \par}}
\newcommand{\URLaddressD}[1]{\ifthenelse{\equal{#1}{}}{}{\par\noindent {$\phantom{\dag}$~\rm URL: }{\tt  #1} \par}}

\newcommand{\Abstract}[1]{\vspace{6mm}\par\noindent\hspace*{10mm}
\parbox{140mm}{\small {\bf Abstract.} #1}\par}
\newcommand{\Keywords}[1]{\vspace{3mm}\par\noindent\hspace*{10mm}
\parbox{140mm}{\small {\bf Key words:} \rm #1}\par}
\newcommand{\Classification}[1]{\vspace{3mm}\par\noindent\hspace*{10mm}
\parbox{140mm}{\small {\it 2000 Mathematics Subject Classification:} \rm #1}\vspace{3mm}\par}
\newcommand{\ShortArticleName}[1]{\renewcommand{\oddhead}{#1}}
\newcommand{\AuthorNameForHeading}[1]{\renewcommand{\evenhead}{#1}}

\long\def\@makecaption#1#2{%\vskip\abovecaptionskip
  \sbox\@tempboxa{\small \textbf{#1.}\ \ #2}%
  \ifdim \wd\@tempboxa >\hsize
    {\small \textbf{#1.}\ \ #2}\par \else
    \global \@minipagefalse
    \hb@xt@\hsize{\hfil\box\@tempboxa\hfil}%
  \fi \vskip\belowcaptionskip}
\def\numberwithin#1#2{\@ifundefined{c@#1}{\@nocounterr{#1}}{%
  \@ifundefined{c@#2}{\@nocnterr{#2}}{%
  \@addtoreset{#1}{#2}%
  \toks@\@xp\@xp\@xp{\csname the#1\endcsname}%
  \@xp\xdef\csname the#1\endcsname
    {\@xp\@nx\csname the#2\endcsname.\the\toks@}}}}
\def\E^#1{{\buildrel #1 \over\vee}}
\newtheorem{theorem}{Theorem}

{\theoremstyle{definition}

}

\begin{document}

\FirstPageHeading{V.I. Gerasimenko, Yu.Yu. Fedchun}

\ShortArticleName{On semigroups of large particle systems}

\AuthorNameForHeading{V.I. Gerasimenko}

\ArticleName{On Semigroups of Large Particle Systems\\ and their Scaling Asymptotic Behavior}

\Author{V.I. Gerasimenko$^\ast$\footnote{E-mail: \emph{gerasym@imath.kiev.ua}}
        and Yu.Yu. Fedchun$^\ast$$^\ast$\footnote{E-mail: \emph{fedchun\_yu@ukr.net}}}

\Address{$^\ast$\hspace*{2mm}Institute of Mathematics of NAS of Ukraine,\\
    \hspace*{4mm}3, Tereshchenkivs'ka Str.,\\
    \hspace*{4mm}01601, Kyiv-4, Ukraine}

\Address{$^\ast$$^\ast$Taras Shevchenko National University of Kyiv,\\
    \hspace*{4mm}Department of Mechanics and Mathematics,\\
    \hspace*{4mm}2, Academician Glushkov Av.,\\
    \hspace*{4mm}03187, Kyiv, Ukraine}

\bigskip

\Abstract{We consider semigroups of operators for hierarchies of evolution equations of
large particle systems, namely, of the dual BBGKY hierarchy for marginal observables and
the BBGKY hierarchy for marginal distribution functions. We establish that the generating
operators of the expansions for one-parametric families of operators of these hierarchies
are the corresponding order cumulants (semi-invariants) of semigroups for the Liouville
equations. We also apply constructed semigroups to the description of the kinetic evolution
of interacting stochastic Markovian processes, modeling the microscopic evolution of soft
active matter. For this purpose we consider the mean field asymptotic behavior of the
semigroup generated by the dual BBGKY hierarchy for marginal observables. The constructed
scaling limit is governed by the set of recurrence evolution equations, namely, by the
Vlasov-type dual hierarchy. Moreover, the relationships of this hierarchy of evolution
equations with the Vlasov-type kinetic equation with initial correlations are established.}

\bigskip

\vspace{2pc}
\Classification{35Q83; 47D06; 82B40.}

\makeatletter
\renewcommand{\@evenhead}{
\hspace*{-3pt}\raisebox{-7pt}[\headheight][0pt]{\vbox{\hbox to \textwidth {\thepage \hfil \evenhead}\vskip4pt \hrule}}}
\renewcommand{\@oddhead}{
\hspace*{-3pt}\raisebox{-7pt}[\headheight][0pt]{\vbox{\hbox to \textwidth {\oddhead \hfil \thepage}\vskip4pt\hrule}}}
\renewcommand{\@evenfoot}{}
\renewcommand{\@oddfoot}{}
\makeatother

\newpage
\vphantom{math}

\protect\tableofcontents
%\newpage
\vspace{0.7cm}

\section{Introduction}
The theory of semigroups of linear operators is a powerful and effective tool to study
problems, which arise in the theory of evolution equations of large particle systems
~\cite{BanArl},~\cite{BM}, in particular, semigroups concerned with such a fundamental
problem as the rigorous derivation of kinetic equations ~\cite{CGP97},~\cite{SR12},~\cite{G}.

In the article ~\cite{GF12} we presented the rigorous results on the evolution equations in
functional derivatives for generating functionals of states and observables of classical
large particle systems, namely, the BBGKY (Bogolyubov-Born-Green-Yovan) hierarchy and the
dual BBGKY hierarchy in functional derivatives, respectively. On the basis of the developed
approach nonperturbative solutions of the Cauchy problem of the corresponding hierarchies
of evolution equations were constructed. One of the purposes of present paper is devoted to
discussing the basic properties of semigroups generated by these solutions in suitable Banach
spaces.

In the paper we also consider the asymptotic behavior of the constructed semigroups by the
example of the dynamical system suggested in ~\cite{L11}, which is based on the Markov jump
processes, modeling the microscopic evolution of soft active matter ~\cite{MJRLPRAS},~\cite{ALBA}.
In the work ~\cite{GF13} it was developed an approach to the description of the kinetic evolution of
large number of interacting entities of biological systems within the framework of the evolution
of marginal ($s$-particle) observables ~\cite{BG10}. Using the mean field scaling asymptotics
of the semigroup of the dual BBGKY hierarchy for marginal observables, in the paper we derive
the Vlasov-type kinetic equation with initial correlations that may characterize condensed
states of such systems. We remark that the conventional approach to this problem is based on
the consideration of an asymptotic behavior of a solution of the BBGKY hierarchy for marginal
($s$-particle) distribution functions constructed within the framework of the theory of perturbations
in case of initial data specified by one-particle (marginal) distribution function without correlations,
i.e. such that satisfy a chaos condition ~\cite{CGP97},~\cite{SR12},~\cite{G}.

We outline the structure of the paper and the main results.
Sect.~\ref{sec:2} will be devoted to introducing the basic notions and to discussing the main
properties of semigroups of operators for hierarchies of evolution equations of large particle
systems on suitable Banach spaces.
In the capacity of applications of these results to the description of collective behavior
of large particle systems of soft matter in Sect.~\ref{sec:3} we develop the new method
of the description of the kinetic evolution within the framework of the evolution of observables.
The mean field asymptotic behavior of the semigroup of operators
for the dual BBGKY hierarchy for marginal observables of the Markov jump processes is constructed.
Moreover, the relationships of the limit marginal observables determined by this asymptotics and
the Vlasov-type kinetic equation with initial correlations is established.
Finally in Sect.~\ref{sec:4} we conclude with some observations and remarks.

\section{Semigroup theory of large particle systems}
\label{sec:2}
We introduce the semigroups of operators for hierarchies of evolution equations of large
classical particle systems and consider their main properties in suitable Banach spaces.
The concept of cumulants (semi-invariants) of semigroups of operators is introduced and
it is established that the corresponding order cumulants of semigroups for the Liouville
equations are the generating operators of expansions for a one-parameter families of
operators for such hierarchies of equations.

\subsection{Preliminaries}
\label{sec:21}
A system of a non-fixed number of particles of unit mass is considered ~\cite{CGP97}.
Every $i$-th particle is characterized by its configuration $q_i\in\mathbb{R}^3$ and
momentum $p_i\in\mathbb{R}^3$ variables and we denote by
$x_i\equiv(q_i,p_i)\in\mathbb{R}^3\times\mathbb{R}^3$ its phase coordinates.
Let $C_\gamma$ be the space of sequences $b=(b_0,b_1,\ldots,b_n,\ldots)$ of measurable
bounded functions $b_n\in C_n$, that are symmetric with respect to permutations of the
arguments $b_n(x_1,\ldots,x_n)$ and equipped with the norm
\begin{eqnarray*}
    &&\|b\|_{C_\gamma}=\max_{n\geq0}\frac{\gamma^n}{n!}\|b_n\|_{C_n}=
      \max_{n\geq0}\frac{\gamma^n}{n!}\max_{x_1,\ldots,x_n}
      \big|b_n(x_1,\ldots,x_n)\big|,
\end{eqnarray*}
where $\gamma<1$ is a parameter.

Dynamics of many-particle systems is described by the one-parameter mapping:
$\mathbb{R}\ni t\mapsto S(t)=\oplus_{n=0}^\infty S_n(t)$, defined on the space
$C_\gamma$ as follows
\begin{eqnarray}\label{S}
   &&(S(t)b)_n(x_1,\ldots,x_n)=S_n(t,1,\ldots,n)b_n(x_1,\ldots,x_n)\doteq\\
   &&\hskip+8mm\doteq b_n(X_1(t,x_1,\ldots,x_n),\ldots,X_n(t,x_1,\ldots,x_n)),\nonumber
\end{eqnarray}
where the functions $\{X_i(t,x_1,\ldots,x_n)\}_{i=1}^n$ are solutions of the Cauchy problem
of the Hamilton equations with initial data $x_1,\ldots,x_n$. The Hamiltonian of the $n$-particle
system has the form
\begin{eqnarray}\label{ham1}
  &&H_n=\sum_{i=1}^n K(p_i)+\sum_{i<j=1}^n\Phi(q_i-q_j),
\end{eqnarray}
where $K(p_i)$ is a kinetic energy of the $i$-th particle and $\Phi(q_i-q_j)$ is a two-body
interaction potential. Hereafter we will assume that the function $\Phi$ satisfies conditions
which guarantee the existence of a global in time solution of the Hamilton equations for
finitely many particles ~\cite{CGP97}.

On the space $C_n$ one-parameter mapping (\ref{S}) is an isometric $\ast$-weak
continuous group of operators. The infinitesimal generator $\mathcal{L}_{n}$ of this group
of operators is a closed operator for the $\ast$-weak topology, and on its domain of the
definition $\mathcal{D}(\mathcal{L}_n)\subset C_n$ it is defined in the sense of the
$\ast$-weak convergence of the space $C_n$ by the operator
\begin{eqnarray}\label{infOper1}
    &&\lim\limits_{t\rightarrow 0}\frac{1}{t}\big(S_n(t)b_n-b_n \big)
        =\{b_n, H_n\}\doteq\mathcal{L}_n b_n,
\end{eqnarray}
where the function $H_{n}$ is the Hamilton function (\ref{ham1}) and the symbol $\{\cdot,\cdot\}$
denotes the Poisson brackets. The Liouville operator $\mathcal{L}_{n}$ has the structure
\begin{eqnarray}\label{L}
    &&\mathcal{L}_n=\sum\limits_{j=1}^{n}\mathcal{L}(j)+
        \sum\limits_{j_{1}\neq j_{2}=1}^{n}\mathcal{L}_{\mathrm{int}}(j_{1},j_{2}),
\end{eqnarray}
where the operator $\mathcal{L}(j)\doteq\langle p_j,\frac{\partial}{\partial q_j}\rangle$
is defined on the set $C_{n,0}\subset C_{n}$ of continuously differentiable functions with
compact supports, and on functions $b_n\in C_{n,0}\subset C_{n}$
the operator $\mathcal{L}_{\mathrm{int}}(j_{1},j_{2})$ is given by the formula
\begin{eqnarray}\label{L_int}
    &&\mathcal{L}_{\mathrm{int}}(j_{1},j_{2})b_n=
       -\langle\frac{\partial}{\partial q_{j_{1}}}\Phi(q_{j_{1}}-q_{j_{2}}),
       \frac{\partial}{\partial p_{j_{1}}}\rangle b_n.
\end{eqnarray}
In (\ref{L_int}) the symbol $\langle \cdot,\cdot \rangle$ means a scalar product.
The Liouville operator $\mathcal{L}_{n}$ is defined in paper ~\cite{G} in case of
a system with hard spheres collisions.

Thus, if $A^0\in C_{\gamma}$, then the group of operators $S(t)=\oplus_{n=0}^\infty S_n(t)$,
determines a unique solution $A(t)=S(t)A^0$ of the Liouville equation for observables.
For finite sequences of continuously differentiable functions with compact supports
$A^0\in C_{\gamma,0}\subset C_{\gamma}$ it is a classical solution, and for arbitrary
initial data $A^0\in C_{\gamma}$ it is a generalized solution of the Liouville equation.

The average values of observables (mean values of observables) are defined by the positive
continuous linear functional on the space $C_{\gamma}$
\begin{eqnarray}\label{af}
     &&(A(t),D^0)\doteq(I,D^0)^{-1}\sum\limits_{n=0}^{\infty}\frac{1}{n!}
         \,\int_{(\mathbb{R}^{3}\times\mathbb{R}^{3})^n}dx_{1}\ldots dx_{n}\,A_{n}(t)\,D_{n}^0,
\end{eqnarray}
where $D^0=(1,D_{1}^0,\ldots,D_{n}^0,\ldots)$ is a sequence of initial distribution functions
defined on the corresponding phase spaces that describes the state of a system of a non-fixed
number of particles \cite{CGP97} and $(I,D^0)={\sum\limits}_{n=0}^{\infty}\frac{1}{n!}
\int_{(\mathbb{R}^{3}\times\mathbb{R}^{3})^n}dx_{1}\ldots dx_{n}D_{n}^0$ is a normalizing factor
(the grand canonical partition function).

Let $L_{\alpha}^{1}=\oplus_{n=0}^\infty {\alpha}^{n}L_{n}^{1}$ be the space of sequences
of integrable functions $f_n\in L_{n}^{1}$, that are symmetric with
respect to permutations of the arguments $f_n(x_1,\ldots,x_n)$ and equipped with the norm
\begin{eqnarray*}
    &&\|f\|_{L_{\alpha}^{1}}=\sum_{n=0}^\infty\alpha^{n}\|f_n\|_{L_{n}^{1}}=
      \sum_{n=0}^\infty\alpha^{n}\int_{(\mathbb{R}^{3}\times\mathbb{R}^{3})^n}dx_{1}\ldots dx_{n}
      \big|f_n(x_1,\ldots,x_n)\big|,
\end{eqnarray*}
where $\alpha>1$ is a parameter. Then for $D^0\in L_{\alpha}^{1}$ and $A(t)\in C_{\gamma}$
functional (\ref{af}) exists and determines a duality between observables and states.

We define dual group of operators $S^{\ast}(t)$ to the group of operators (\ref{S})
in the sense of the bilinear form of mean values (\ref{af}), i.e. $(S(t)b,f)=(b,S^{\ast}(t)f)$.
For the dual group of operators $S^{\ast}(t)$ the following equality is true:
\begin{eqnarray}\label{S*}
   &&(S^{\ast}(t)f)_n(x_1,\ldots,x_n)=S^{\ast}_n(t)f_n(x_1,\ldots,x_n)=\\
   &&\hskip+8mm=S_n(-t)f_n(x_1,\ldots,x_n),\nonumber
\end{eqnarray}
where the operator $S_n(-t)$ is defined by formula (\ref{S}).

On the space $L_{n}^{1}$ one-parameter mapping (\ref{S*}) is an isometric strong
continuous group of operators. The infinitesimal generator $\mathcal{L}^\ast_{n}$
of this group of operators is a closed operator and on its domain of the
definition $\mathcal{D}(\mathcal{L}^\ast_n)\subset L_{n}^{1}$ it is defined in the
sense of the norm convergence of the space $L_{n}^{1}$ by the operator
\begin{eqnarray*}
    &&\lim\limits_{t\rightarrow 0}\big\|\frac{1}{t}(S^\ast_n(t)f_n-f_n)
       -\mathcal{L}^\ast_nf_n\big\|_{L_{n}^{1}}=0,
\end{eqnarray*}
where the operator $\mathcal{L}^\ast$ is an adjoint operator to the Liouville operator (\ref{infOper1})
in the sense of the bilinear form of mean values (\ref{af}), and the following equality is true:
\begin{eqnarray*}
    &&\mathcal{L}^\ast_n f_n=-\mathcal{L}_n f_n.
\end{eqnarray*}

Thus, if $D^0\in L_{\alpha}^{1}$, then the group of operators $S^\ast(t)=\oplus_{n=0}^\infty S^\ast_n(t)$,
determines a unique solution $D(t)=S^\ast(t)D^0$ of the Liouville equation for states.
For finite sequences of continuously differentiable functions with compact supports
$D^0\in L_{\alpha,0}^{1}\subset L_{\alpha}^{1}$ it is a strong solution, and for arbitrary
initial data $D^0\in L_{\alpha}^{1}$ it is a weak solution of the Liouville equation.

\subsection{Cumulants of semiroups of operators}
\label{sec:22}
For systems of a finite average number of particles there exists an equivalent method to
describe the evolution of observables and states in comparison with considered above, namely,
within the framework of semigroups for marginal observables and marginal distribution functions.
In suitable Banach spaces such semigroups give an opportunity to describe also the evolution
of infinitely many particles.

First of all we shall introduce the notion of cumulants (semi-invariants) of semigroups (\ref{S})
for the Liouville equations which are the generating operators of expansions for one-parameter
families of operators for hierarchies of evolution equations for marginal observables and
marginal distribution functions.

Let us introduce some abridged notations: $Y\equiv(1,\ldots,s),\, X\equiv(j_1,\ldots,j_{n})\subset Y$
and $\{Y\setminus X\}$ is the set, consisting of one element $Y\setminus X=(1,\ldots,s)\setminus(j_1,\ldots,j_{n})$.
The $(1+n)th$-order cumulant of groups of operators (\ref{S}) is defined by the following expansion ~\cite{GF12}:
\begin{eqnarray}\label{cumulant}
    &&\hskip-8mm\mathfrak{A}_{1+n}(t,\{Y\setminus X\},\,X)\doteq
       \sum\limits_{\mathrm{P}:\,(\{Y\setminus X\},\,X)={\bigcup}_i X_i}
       (-1)^{\mathrm{|P|}-1}({\mathrm{|P|}-1})!
       \prod_{X_i\subset \mathrm{P}}S_{|\theta(X_i)|}(t,\theta(X_i)),\quad n\geq0,
\end{eqnarray}
where the symbol ${\sum}_\mathrm{P}$ is a sum over all possible partitions $\mathrm{P}$
of the set $(\{Y\setminus X\},j_1,\ldots,j_{n})$ into $|\mathrm{P}|$ nonempty
mutually disjoint subsets $ X_i\subset(\{Y\setminus X\},X)$ and $\theta(\cdot)$ is the
declusterization mapping defined as follows: $\theta(\{Y\setminus X\},\,X)=Y$. For example,
\begin{eqnarray*}
    &&\mathfrak{A}_{1}(t,\{Y\})=S_{s}(t,Y),\\
    &&\mathfrak{A}_{2}(t,\{Y\setminus (j)\},j)=S_{s}(t,Y)
       -S_{s-1}(t,Y\setminus(j))S_{1}(t,j).\nonumber
\end{eqnarray*}

Let us indicate some properties of cumulants (\ref{cumulant}). If $n=0$, for
$b_{s}\in C_{s,0}\subset C_{\gamma}$ in the sense of
the $\ast$-weak convergence of the space $C_{s}$ the generator of
first-order cumulant (\ref{cumulant}) is given by the Liouville operator (\ref{L})
\begin{eqnarray}\label{1}
    &&\lim\limits_{t\rightarrow 0}\frac{1}{t}(\mathfrak{A}_{1}(t,\{Y\})-I)b_{s}=
       \mathcal{L}_{s}b_{s}.
\end{eqnarray}
In case of $n=1$ for $b_{s}\in C_{s,0}\subset C_{\gamma}$
we obtain the following equality in the sense of the $\ast$-weak convergence of the space $C_{s}$
\begin{eqnarray}\label{2}
   &&\lim\limits_{t\rightarrow 0}\frac{1}{t}\,\mathfrak{A}_{2}(t,\{Y\setminus(j)\},j)b_{s}
       =\sum_{i\in(Y\setminus(j))}\mathcal{L}_{\mathrm{int}}(i,j)b_{s},
\end{eqnarray}
where the operator $\mathcal{L}_{\mathrm{int}}(i,j)$ is defined by formula (\ref{L_int}), and
for $n\geq2$ as a consequence of the fact that we consider a system of particles with two-body
interaction potential (\ref{ham1}), it holds
\begin{eqnarray}\label{3}
   &&\lim\limits_{t\rightarrow 0}\frac{1}{t}\,\mathfrak{A}_{1+n}(t,\{Y\setminus X\},\,X)b_{s}=0.
\end{eqnarray}
We remark that in case of $n$-body interaction potentials such derivative is determined by the
corresponding operator similar to the case of the second order cumulant.

If $b_{s}\in C_{s}$, then for $(1+n)th$-order cumulant (\ref{cumulant}) of groups of operators
(\ref{S}) the following estimate is valid:
\begin{eqnarray}\label{estd}
   &&\big\|\mathfrak{A}_{1+n}(t,\{Y\setminus X\},\,X)b_{s}\big\|_{C_{s}}
       \leq \sum\limits_{\mathrm{P}:\,(\{Y\setminus X\},\,X)={\bigcup}_i X_i}
       (|\mathrm{P}|-1)!\big\|b_{s}\big\|_{C_{s}}\leq \\
   &&\hskip+8mm\leq \sum\limits_{k=1}^{n+1}\mathrm{s}(n+1,k)(k-1)!\big\|b_{s}\big\|_{C_{s}}
        \leq n!e^{n+2}\big\|b_{s}\big\|_{C_{s}},\nonumber
\end{eqnarray}
where $\mathrm{s}(n+1,k)$ are the Stirling numbers of the second kind.

\subsection{Group of operators for the dual BBGKY hierarchy}
\label{sec:23}
Let $b\in C_{\gamma}$ and $\gamma<e^{-1}$, then the one-parameter mapping for the dual BBGKY
hierarchy $\mathbb{R}\ni t\mapsto U(t)b$ is defined by the following expansion:
\begin{eqnarray}\label{mapdual}
    &&\hskip-8mm(U(t)b)_{s}(x_1,\ldots,x_s)\doteq\\
    &&\hskip-8mm\doteq\sum_{n=0}^s\,\frac{1}{(s-n)!}\sum_{j_1\neq\ldots\neq j_{s-n}=1}^s
        \mathfrak{A}_{1+n}(t,\{Y\setminus X\},\,X)\,
        b_{s-n}((x_1,\ldots,x_s)\setminus (x_{j_1},\ldots,x_{j_n})),\quad s\geq1,\nonumber
\end{eqnarray}
where the generating operator $\mathfrak{A}_{1+n}(t)$ is the $(1+n)th$-order cumulant
(\ref{cumulant}) of groups of operators (\ref{S}).

The one-parameter mapping for the dual BBGKY hierarchy (\ref{mapdual}) has the following
properties.
\begin{theorem}
If $\gamma<e^{-1}$, then on the space $C_{\gamma}$ one-parameter mapping
(\ref{mapdual}) is a $C_{0}^{\ast}$-group.
The infinitesimal generator $\mathcal{B}={\bigoplus\limits}_{s=0}^{\infty}\mathcal{B}_{s}$
of this group of operators is a closed operator for the $\ast$-weak topology and on the
domain of the definition $\mathcal{D}(\mathcal{B})\subset C_{\gamma}$ which is the everywhere
dense set for the $\ast$-weak topology of the space $C_{\gamma}$ it is defined by the operator
\begin{eqnarray}\label{d}
    &&\hskip-7mm(\mathcal{B}b)_{s}(x_1,\ldots,x_s)\doteq\mathcal{L}_{s}(Y)b_{s}(x_1,\ldots,x_s)+\\
    &&\hskip+8mm +\sum_{j_1\neq j_{2}=1}^s\mathcal{L}_{\mathrm{int}}(j_1,j_{2})
       b_{s-1}((x_1,\ldots,x_s)\setminus (x_{j_1})),\quad s\geq 1, \nonumber
\end{eqnarray}
where the operators $\mathcal{L}_{s}$ and $\mathcal{L}_{\mathrm{int}}$ are given by formulas
(\ref{infOper1}) and (\ref{L_int}), respectively.
\end{theorem}

Indeed, under the condition that $\gamma<e^{-1}$, owing to estimate (\ref{estd}), we have
\begin{eqnarray*}\label{es}
    &&\big\|(U(t)b)\big\|_{C_{\gamma}}
       \leq e^2(1-\gamma e)^{-1}\big\|b\big\|_{C_{\gamma}}.
\end{eqnarray*}
Hence the group of operators (\ref{mapdual}) is defined on the space $C_{\gamma}$.

On the space $C_{\gamma}$ the $\ast$-weak continuity property over the parameter
$t\in \mathbb{R}$ of the group of operators $\{U(t)\}_{t\in\mathbb{R}}$ is a consequence
of the $\ast$-weak continuity of the group of operators (\ref{S}) for the Liouville equation.

To construct an infinitesimal generator of the group $\{U(t)\}_{t\in\mathbb{R}}$ we use
equalities (\ref{1})-(\ref{3}). Then for the group of operators (\ref{mapdual}) on $C_{\gamma,0}$
in the sense of the $\ast$-weak convergence the equality holds
\begin{eqnarray*}
    &&\hskip-8mm\lim\limits_{t\rightarrow 0}\frac{1}{t}\big((U(t)b)_{s}-b_{s}\big)=
       \mathcal{L}_{s}(Y)b_{s}(x_1,\ldots,x_s)+\sum_{j_1\neq j_{2}=1}^s\mathcal{L}_{\mathrm{int}}(j_1,j_{2})
       b_{s-1}((x_1,\ldots,x_s)\setminus (x_{j_1})).
\end{eqnarray*}

We consider the structure of an infinitesimal generator of the group of operators (\ref{mapdual}).
Introducing the operator (an analog of the creation operator ~\cite{BG10}):
\begin{eqnarray}\label{oper_znuw}
   &&(\mathfrak{a}^{+}b)_{s}(x_1,\ldots,x_s)\doteq
        \sum_{j=1}^s\,b_{s-1}(x_1,\ldots,x_{j-1},x_{j+1},\ldots,x_s),
\end{eqnarray}
defined on $C_{\gamma}$, in the general case infinitesimal generator (\ref{d}) is also represented
in the following form:
\begin{eqnarray}\label{gg}
    &&\mathcal{B}=\mathcal{L}+\sum\limits_{n=1}^{\infty}\frac{1}{n!}
     \big[\ldots\big[\mathcal{L},\underbrace{\mathfrak{a}^{+} \big],\ldots,\mathfrak{a}^{+}}_{\hbox{n-times}}\big]=\\
    &&\hskip+7mm=e^{-\mathfrak{a}^{+}}\mathcal{L}e^{\mathfrak{a}^{+}},\nonumber
\end{eqnarray}
where the symbol $\big[\cdot,\cdot\big]$ denotes the commutator of operators.
As a consequence of the fact that we consider a system of particles with
a two-body interaction potential (\ref{ham1}), the following equalities hold:
\begin{eqnarray*}
   &&(\big[\mathcal{L},\mathfrak{a}^{+}\big]b)_s(x_1,\ldots,x_s)=
       \sum_{j_1\neq j_{2}=1}^s\mathcal{L}_{\mathrm{int}}(j_1,j_{2})
       b_{s-1}((x_1,\ldots,x_s)\setminus (x_{j_1})),\\
   &&(\big[\big[\mathcal{L},\mathfrak{a}^{+}\big],\mathfrak{a}^{+}\big]b)_s(x_1,\ldots,x_s)=0.
\end{eqnarray*}

Thus, if $B^0\in C_{\gamma}$, then under the condition that $\gamma<e^{-1}$, the group of operators
(\ref{mapdual}) determines a unique solution $B(t)=U(t)B^0$ of the Cauchy problem of the dual
BBGKY hierarchy for marginal observables ~\cite{BG10}. For $B^0\in C_{\gamma}^0\subset C_{\gamma}$
it is a classical solution, and for arbitrary initial data $B^0\in C_{\gamma}$ it is a generalized
solution.

We remark that, since hierarchy of evolution equations for marginal observables has the structure
of recurrence equations, we also deduce that the solution expansion can be also constructed by
successive integration of the inhomogeneous Liouville equations. The solution $B(t)=U(t)B^0$ may
be represented in the form of the perturbation (iteration) series as a result of applying of analogs
of the Duhamel equation ~\cite{BanArl} to cumulants (\ref{cumulant}) of groups of operators (\ref{S}).

We emphasize that the evolution of marginal observables of both finitely and infinitely many
particles is described by the Cauchy problem of the dual BBGKY hierarchy. For finitely many
particles the dual BBGKY hierarchy is equivalent to the Liuville equations for observables.

\subsection{Group of operators for the BBGKY hierarchy}
\label{sec:24}
We define the dual group of operators $\{U^{\ast}(t)\}_{t\in\mathbb{R}}$ to the group of operators
(\ref{mapdual}) in the sense of the bilinear form of mean values (\ref{af}),
i.e. $(U(t)b,f)=(b,U^{\ast}(t)f)$.

Let $f\in L_{\alpha}^{1}$ and $\alpha>e$, then a one-parameter mapping for the BBGKY hierarchy
is defined by the following series expansion:
\begin{eqnarray}\label{RozvBBGKY}
  &&\hskip-5mm\mathbb{R}\ni t\mapsto(U^{\ast}(t)f)_{s}(x_1,\ldots,x_{s})\doteq\\
  &&\hskip-5mm\doteq\sum\limits_{n=0}^{\infty}\frac{1}{n!}
      \int_{(\mathbb{R}^{3}\times\mathbb{R}^{3})^n}dx_{s+1}\ldots dx_{s+n}
      \mathfrak{A}_{1+n}(t,\{Y\},X\setminus Y)f_{s+n}(x_1,\ldots,x_{s+n}),\nonumber
\end{eqnarray}
where the generating operator $\mathfrak{A}_{1+n}(t,\{Y\},\,X\setminus Y)$ is the $(1+n)th$-order
cumulant of groups (\ref{S*}):
\begin{eqnarray}\label{cum}
   &&\hskip-9mm\mathfrak{A}_{1+n}(t,\{Y\},\,X\setminus Y)\doteq
        \sum\limits_{\mathrm{P}:\,(\{Y\},\,X\setminus Y)={\bigcup}_i X_i}(-1)^{|\mathrm{P}|-1}
       (|\mathrm{P}|-1)!\prod_{X_i\subset \mathrm{P}}S^{\ast}_{\theta(|X_i|)}(t,\theta(X_i)),\quad n\geq0.
\end{eqnarray}
In expansion (\ref{cum}) we used notations accepted above and the symbol ${\sum}_\mathrm{P}$ is the
sum over all possible partitions $\mathrm{P}$ of the set $(\{Y\},\,X\setminus Y)=(\{Y\},s+1,\ldots,s+n)$
into $|\mathrm{P}|$ nonempty mutually disjoint subsets $ X_i\subset(\{Y\},\,X\setminus Y)$.

The one-parameter mapping for the BBGKY hierarchy (\ref{RozvBBGKY}) has the following properties.
\begin{theorem}
If $\alpha>e$, then on the space $L^{1}_{\alpha}$ one-parameter mapping (\ref{RozvBBGKY})
is a $C_{0}$-group.
On the subspace $L^{1}_{\alpha,0}\subset L^{1}_{\alpha}$ the infinitesimal generator
$\mathcal{B}^{\ast}={\bigoplus\limits}_{n=0}^{\infty}\mathcal{B}^{\ast}_{n}$
of the group of operators (\ref{RozvBBGKY}) is defined by the operator
\begin{eqnarray}\label{big}
   &&\hskip-7mm (\mathcal{B}^{\ast}f)_{s}(x_1,\ldots,x_{s})=
         \mathcal{L}^{\ast}_{s}f_{s}(x_1,\ldots,x_{s})+\\
   &&\hskip+8mm+\sum\limits_{i=1}^{s}\int_{\mathbb{R}^3\times\mathbb{R}^3}d x_{s+1}
         \mathcal{L}^{\ast}_{\mathrm{int}}(i,s+1)f_{s+1}(x_1,\ldots,x_{s+1}),\quad s\geq 1,\nonumber
\end{eqnarray}
where on the subspace $L_{s,0}^{1}\subset L^{1}_{s}$ the operator $\mathcal{L}^{\ast}_{s}$ is an adjoint
operator to the Liouville operator (\ref{infOper1}) and $\mathcal{L}_{s}^{\ast}f_{s}=-\mathcal{L}_{s}f_{s}$.
\end{theorem}

One-parameter mapping (\ref{RozvBBGKY}) is defined on the space $L_{\alpha}^{1}$ provided that $\alpha>e$
and, according to inequality (\ref{estd}), the estimate holds
\begin{eqnarray*}\label{estim_L_{1}}
       &&\|U(t)f\|_{L_{\alpha}^{1}}\leq e^{2}(1-\frac{e}{\alpha})^{-1}\|f\|_{L_{\alpha}^{1}}.
\end{eqnarray*}

The property of the strong continuity of the group $\{U^{\ast}(t)\}_{t\in\mathbb{R}}$ over the parameter
$t\in\mathbb{R}$ is a consequence of the strong continuity of group (\ref{S*}) for the Liouville equation.

To construct an infinitesimal generator of this group of operators (\ref{RozvBBGKY}),
we take into account the validity for $f\in L^{1}_{\alpha,0}$ the following equalities:
\begin{eqnarray*}\label{deriv}
 &&\lim_{t\rightarrow 0}\big\|\frac{1}{t}\big(\mathfrak{A}_{1}(t,\{Y\})-I\big)f_s
          -\mathcal{L}^{\ast}(Y)f_s\big\|_{L^{1}_{s}}=0,\\
 &&\lim\limits_{t\rightarrow 0}\big\|\frac{1}{t}\,\mathfrak{A}_{2}(t,\{Y\},s+1)f_{s+1}
       -\sum_{j=1}^{s}\mathcal{L}^{\ast}_{\mathrm{int}}(j,s+1)f_{s+1}\big\|_{L^{1}_{s+1}}=0,\\
  &&\lim\limits_{t\rightarrow 0}\big\|\frac{1}{t}\,
   \mathfrak{A}_{1+n}(t,\{Y\setminus X\},X)f_{s+n}\big\|_{L^{1}_{s+n}}=0.
\end{eqnarray*}
Then for the group of operators (\ref{RozvBBGKY}) we finally derive in the sense of the norm convergence
\begin{eqnarray*}
   &&\lim_{t\rightarrow 0}\big\|\frac{1}{t}\big(U(t)f-f\big)
          -\mathcal{B}^{\ast}f\big\|_{L^{1}_{\alpha}}=0,
\end{eqnarray*}
where the operator $\mathcal{B}^{\ast}$ on $L^{1}_{\alpha,0}$ is given by formula (\ref{big}).

In terms of the operator adjoint to operator (\ref{oper_znuw}) (an analog of the annihilation operator)
defined on the space $L^{1}_{\alpha}$ by the formula
\begin{eqnarray}\label{oper_an}
         \big(\mathfrak{a}f\big)_{s}(x_1,\ldots,x_{s})\doteq\int dx_{s+1}f_{s+1}(x_1,\ldots,x_{s+1}),
\end{eqnarray}
an infinitesimal generator of the group of operators (\ref{RozvBBGKY}) is represented in the following
form (the adjoint operator to operator (\ref{gg})):
\begin{eqnarray}\label{rbg}
    &&\mathcal{B}^{\ast}=\mathcal{L}^{\ast}+ \sum\limits_{n=1}^{\infty}\frac{1}{n!}
     \big[\underbrace{\mathfrak{a},\ldots,\big[{\mathfrak{a}}}_{\hbox{n-times}},\mathcal{L}^{\ast}\big]\ldots \big]=\\
    &&\hskip+7mm=e^{\mathfrak{a}}\mathcal{L}^{\ast}e^{-\mathfrak{a}}.\nonumber
\end{eqnarray}
Representation (\ref{rbg}) is valid in consequence of definition (\ref{oper_an})
and the validity for a system of particles interacting through a two-body potential of
the equalities:
\begin{eqnarray*}
   &&\big(\big[\mathfrak{a},\mathcal{L}^{\ast}\big]f\big)_s(x_1,\ldots,x_s)=
       \sum\limits_{i=1}^{s}\int dx_{s+1}\mathcal{L}_{\mathrm{int}}^{\ast}(i,s+1)f_{s+1}(x_1,\ldots,x_{s+1}),\\
   &&\big(\big[\mathfrak{a},\big[\mathfrak{a},\mathcal{L}^{\ast}\big]\big]f\big)_s(x_1,\ldots,x_s)=0.
\end{eqnarray*}

The evolution of all possible states both finitely and infinitely many particles is described
by the Cauchy problem of the BBGKY hierarchy for marginal distribution functions.
If $F^0\in L^{1}_{\alpha}$, then under the condition that $\alpha>e$, the group of operators
(\ref{RozvBBGKY}) determines unique solution $F(t)=U^{\ast}(t)F^0$ of the Cauchy problem of
the BBGKY hierarchy for marginal distribution functions ~\cite{CGP97}.
For $F^0\in L^{1}_{\alpha,0}\subset L^{1}_{\alpha}$ it is a strong solution, and for arbitrary
initial data $F^0\in L^{1}_{\alpha}$ it is a weak solution.

We note that the properties of semigroups of the hierarchies of evolution equations
of large particle systems  in suitable Banach spaces were considered in paper ~\cite{BG}
for granular gases and in review ~\cite{G12} for quantum gases.

\section{Application to kinetic theory of soft active matter}
\label{sec:3}
As is known the collective behavior of many-particle systems can be effectively described
within the framework of a one-particle (marginal) distribution function governed by the
kinetic equation in a suitable scaling limit of underlying dynamics of states. The conventional
approach to this problem is based on the consideration of an asymptotic behavior of a
solution of the BBGKY hierarchy for marginal ($s$-particle) distribution functions constructed
within the framework of the theory of perturbations in case of initial data specified by
one-particle (marginal) distribution function without correlations, i.e. such that satisfy
a chaos condition ~\cite{CGP97},~\cite{SR12}.

Another method of the description of the many-particle evolution is given within the
framework of marginal ($s$-particle) observables governed by the dual BBGKY hierarchy.
In this section we consider the problem of the rigorous description of the kinetic evolution
within the framework of an asymptotic behavior of marginal ($s$-particle) observables governed
by the dual BBGKY hierarchy. To this end we construct the mean field scaling limit of the
semigroup represented by expansions (\ref{mapdual}) in case of dynamical systems modeling
many-entity biological systems. One of the advantages of such an approach is the
possibility to construct the kinetic equations in scaling limits, involving correlations
at initial time, that can characterize the condensed states of biological systems.

\subsection{Stochastic dynamics of many-entity systems}
\label{sec:31}
The many-entity biological systems are dynamical systems displaying a collective
behavior which differs from the statistical behavior of usual gases. To specify
such nature of entities we consider the dynamical system suggested in paper ~\cite{L11},
which is based on the Markov jump processes that can represent the intrinsic properties
of living creatures (soft active matter ~\cite{MJRLPRAS}).

We consider a system of entities of various $M$ subpopulations introduced in paper ~\cite{L11}
in case of non-fixed, i.e. arbitrary, but finite average number of entities. Every $i$-th
entity is characterized by: $\textbf{u}_i=(j_i,u_i)\in\mathcal{J}\times\mathcal{U}$, where
$j_i\in\mathcal{J}\equiv(1,\ldots,M)$ is a number of its subpopulation, and
$u_i\in\mathcal{U}\subset\mathbb{R}^{d}$ is its microscopic state ~\cite{L11}.
The stochastic dynamics of entities of various subpopulations is described by the semigroup
$e^{t\Lambda}=\oplus_{n=0}^\infty e^{t\Lambda_n}$ of the Markov jump process defined on the
space $C_\gamma$ of sequences  $b=(b_0,b_1,\ldots,b_n,\ldots)$ of measurable bounded
functions $b_n(\textbf{u}_1,\ldots,\textbf{u}_n)$ that are symmetric with respect to
permutations of the arguments $\textbf{u}_1,\ldots,\textbf{u}_n$ and equipped with the norm:
\begin{eqnarray*}
    &&\|b\|_{C_\gamma}=\max_{n\geq0}\frac{\gamma^n}{n!}\|b_n\|_{C_n}=
      \max_{n\geq0}\frac{\gamma^n}{n!}\max_{j_1,\ldots,j_n}\max_{u_1,\ldots,u_n}
      \big|b_n(\textbf{u}_1,\ldots,\textbf{u}_n)\big|,
\end{eqnarray*}
where $\gamma<1$ is a parameter. The generator $\Lambda_n$ (the Liouville operator of $n$ entities)
is defined on the subspace $C_n$ of the space $C_\gamma$ and it has the following structure ~\cite{L11}:
\begin{eqnarray}\label{gen_obs_gener}
    &&\hskip-7mm(\Lambda_n b_n)(\textbf{u}_1,\ldots,\textbf{u}_n)\doteq
        \sum_{m=1}^M \varepsilon^{m-1}\sum_{i_1\neq\ldots\neq i_m=1}^n
        (\Lambda^{[m]}(i_1,\ldots,i_m)b_n)(\textbf{u}_1,\ldots,\textbf{u}_n)=\\
    &&=\sum_{m=1}^M \varepsilon^{m-1}\sum_{i_1\neq\ldots\neq i_m=1}^n
        a^{[m]}(\textbf{u}_{i_1},\ldots,\textbf{u}_{i_m})\big(\int_{\mathcal{J}\times\mathcal{U}}
        A^{[m]}(\textbf{v};\textbf{u}_{i_1},\ldots,\textbf{u}_{i_m})\times\nonumber\\
    &&\hskip+7mm\times b_n(\textbf{u}_1,\ldots,\textbf{u}_{i_1-1},\textbf{v},\textbf{u}_{i_1+1},
        \ldots\textbf{u}_n)d\textbf{v}-b_n(\textbf{u}_1,\ldots,\textbf{u}_n)\big)\nonumber,
\end{eqnarray}
where $\varepsilon>0$ is a scaling parameter, the functions
$a^{[m]}(\textbf{u}_{i_1},\ldots,\textbf{u}_{i_m}),\,m\geq1,$ characterize the
interaction between entities, in particular, in case of $m=1$ it is the interaction
of entities with an external environment. These functions are measurable positive bounded
functions on $(\mathcal{J}\times\mathcal{U})^n$ such that:
$0\leq a^{[m]}(\textbf{u}_{i_1},\ldots,\textbf{u}_{i_m})\leq a^{[m]}_*,$
where $a^{[m]}_*$ is some constant. The functions
$A^{[m]}(\textbf{v};\textbf{u}_{i_1},\ldots,\textbf{u}_{i_m}),\,m\geq1$, are measurable positive
integrable functions which describe the probability of the transition of the $i_1$ entity in the
microscopic state $u_{i_1}$ to the state $v$ as a result of the interaction with entities in the
states $u_{i_2},\ldots,u_{i_m}$ (in case of $m=1$ it is the interaction with an external environment).
The functions $A^{[m]}(\textbf{v};\textbf{u}_{i_1},\ldots,\textbf{u}_{i_m}),\,m\geq1$, satisfy
the conditions:
$\int_{\mathcal{J}\times\mathcal{U}}A^{[m]}(\textbf{v};\textbf{u}_{i_1},\ldots,\textbf{u}_{i_m})d\textbf{v}=1$.
We refer to paper ~\cite{L11}, where examples of the functions $a^{[m]}$ and $A^{[m]}$ are given
in the context of biological systems.

In case of $m=1$ generator (\ref{gen_obs_gener}) has the form $\sum_{i_1=1}^n\Lambda^{[1]}_n(i_1)$
and it describes the free stochastic evolution of entities. The case of $m\geq2$ corresponds to a
system with the $m$-body interaction of entities in the sense accepted in kinetic theory.
Further we restrict ourself by the case of a two-body interaction, i.e. $M=2$.

On the space $C_n$ the one-parameter mapping $\{e^{t\Lambda_{n}}\}_{t\in\mathbb{R}}$ is a bounded $\ast$-weak
continuous semigroup of operators ~\cite{BanArl}.

Thus, for a system under consideration semigroup (\ref{mapdual}) for marginal observables
is represented by the following expansion:
\begin{eqnarray}\label{sdh}
   &&\hskip-7mm (U(t)b)_{s}(t,\textbf{u}_1,\ldots,\textbf{u}_s)=
      \sum_{n=0}^{s-1}\,\frac{1}{n!}\sum_{j_1\neq\ldots\neq j_{n}=1}^s\mathfrak{A}_{1+n}(t,\{Y\setminus X\},\,X)\,
      b_{s-n}(\textbf{u}_1,\ldots,\\
   &&\hskip+25mm\textbf{u}_{j_1-1},\textbf{u}_{j_1+1},\ldots,\textbf{u}_{j_n-1},
      \textbf{u}_{j_n+1},\ldots,\textbf{u}_s), \quad s\geq 1,\nonumber
\end{eqnarray}
where the $(1+n)th$-order cumulant of the semigroups $\{e^{t\Lambda_{k}}\}_{t\in\mathbb{R}},\, k\geq1,$
is determined by the formula
\begin{eqnarray}\label{cumulantd}
   &&\hskip-5mm\mathfrak{A}_{1+n}(t,\{Y\setminus X\},\,X)\doteq
       \sum\limits_{\mathrm{P}:\,(\{Y\setminus X\},\,X)={\bigcup}_i X_i}
       (-1)^{\mathrm{|P|}-1}({\mathrm{|P|}-1})!\prod_{X_i\subset \mathrm{P}}e^{t\Lambda_{|\theta(X_i)|}},
\end{eqnarray}
and we used notations accepted above in (\ref{cumulant}).

The sequence $B(t)=U(t)B^0$ of marginal observables determined by semigroup (\ref{sdh}) is a classical
nonperturbative solution of the Cauchy problem of the dual BBGKY hierarchy for entities ~\cite{GF14}.

\subsection{The mean field limit of the semigroup for the dual BBGKY hierarchy}
\label{sec:33}
We consider the mean field scaling limit of the semigroup given by expansions (\ref{sdh}).

\begin{theorem}
Let $b\in C_\gamma$, then for arbitrary finite time interval there exists the mean field limit
of semigroup represented by expansions (\ref{sdh}) in the sense of the $\ast$-weak convergence
of the space $C_s$
\begin{eqnarray}\label{Iterd}
   &&\hskip-9mm \mathrm{w^{\ast}-}\lim\limits_{\epsilon\rightarrow 0}
       \Big(\sum_{n=0}^{s-1}\,\frac{1}{n!}\sum_{j_1\neq\ldots\neq j_{n}=1}^s
       \epsilon^{-n}\mathfrak{A}_{1+n}\big(t,\{(1,\ldots,s)\setminus(j_1,\ldots,j_{n})\},j_1,\ldots,j_{n}\big)-\\
   &&-\sum\limits_{n=0}^{s-1}\,\int_0^tdt_{1}\ldots\int_0^{t_{n-1}}dt_{n}
      \,e^{(t-t_{1})\sum\limits_{k_{1}=1}^{s}\Lambda^{[1]}(k_{1})}\sum\limits_{i_{1}\neq j_{1}=1}^{s}
      \Lambda^{[2]}(i_{1},j_{1})e^{(t_{1}-t_{2})\sum\limits_{l_{1}=1,l_{1}\neq j_{1}}^{s}\Lambda^{[1]}(l_{1})}\ldots\nonumber\\
   &&\ldots e^{(t_{n-1}-t_{n})\hskip-1mm\sum\limits^{s}_{\mbox{\scriptsize $\begin{array}{c}k_{n}=1,\\
      k_{n}\neq (j_{1},\ldots,j_{n-1})\end{array}$}}\Lambda^{[1]}(k_{n})}
      \sum\limits^{s}_{\mbox{\scriptsize $\begin{array}{c}i_{n}\neq j_{n}=1,\\
      i_{n},j_{n}\neq (j_{1},\ldots,j_{n-1})\end{array}$}}\Lambda^{[2]}(i_{n},j_{n})\times\nonumber\\
   &&\times e^{t_{n}\sum\limits^{s}_{\mbox{\scriptsize $\begin{array}{c}l_{n}=1,\\
      l_{n}\neq (j_{1},\ldots,j_{n})\end{array}$}}\hskip-1mm\Lambda^{[1]}(l_{n})}\Big)
      b_{s-n}^0((\textbf{u}_1,\ldots,\textbf{u}_s)\setminus (\textbf{u}_{j_{1}},\ldots,\textbf{u}_{j_{n}}))=0,\nonumber
\end{eqnarray}
\end{theorem}

The proof of this statement is based on formulas for cumulants (\ref{cumulantd}) of asymptotically
perturbed semigroups of operators  $\{e^{t\Lambda_{k}}\}_{t\in\mathbb{R}},\,\, k\geq2$~\cite{BL14}.

For arbitrary finite time interval the asymptotically perturbed semigroup (\ref{S})
has the following scaling limit in the sense of the $\ast$-weak convergence on the
space $C_s$:
\begin{eqnarray}\label{Kato}
    &&\mathrm{w^{\ast}-}\lim\limits_{\epsilon\rightarrow 0}\big(e^{t\Lambda_{s}}-
       \prod\limits_{j=1}^{s}e^{t\Lambda^{[1]}(j)}\big)b_s=0.
\end{eqnarray}
Taking into account analogs of the Duhamel equations ~\cite{BanArl} for cumulants of asymptotically
perturbed groups of operators (\ref{cumulantd}), in view of formula (\ref{Kato}) we obtain
\begin{eqnarray*}\label{apc}
   &&\hskip-8mm \mathrm{w^{\ast}-}\lim\limits_{\epsilon\rightarrow 0}
       \Big(\epsilon^{-n}\frac{1}{n!}
       \mathfrak{A}_{1+n}\big(t,\{(1,\ldots,s)\setminus(j_1,\ldots,j_{n})\},j_1,\ldots,j_{n}\big)-\\
   &&-\int_0^tdt_{1}\ldots\int_0^{t_{n-1}}dt_{n}
      \,e^{(t-t_{1})\sum\limits_{k_{1}=1}^{s}\Lambda^{[1]}(k_{1})}\sum\limits_{i_{1}\neq j_{1}=1}^{s}
      \Lambda^{[2]}(i_{1},j_{1}) e^{(t_{1}-t_{2})\sum\limits_{l_{1}=1,l_{1}\neq j_{1}}^{s}\Lambda^{[1]}(l_{1})}\ldots \\
   &&\ldots e^{(t_{n-1}-t_{n})\sum\limits^{s}_{\mbox{\scriptsize $\begin{array}{c}k_{n}=1,\\
      k_{n}\neq (j_{1},\ldots,j_{n-1})\end{array}$}}\Lambda^{[1]}(k_{n})}
      \sum\limits^{s}_{\mbox{\scriptsize $\begin{array}{c}i_{n}\neq j_{n}=1,\\
      i_{n},j_{n}\neq (j_{1},\ldots,j_{n-1})\end{array}$}}\Lambda^{[2]}(i_{n},j_{n})\times\\
   &&\times e^{t_{n}\sum\limits^{s}_{\mbox{\scriptsize $\begin{array}{c}l_{n}=1,\\
      l_{n}\neq (j_{1},\ldots,j_{n})\end{array}$}}\Lambda^{[1]}(l_{n})}\Big)
      b_{s-n}((\textbf{u}_1,\ldots,\textbf{u}_s)\setminus (\textbf{u}_{j_{1}},\ldots,\textbf{u}_{j_{n}}))=0,
\end{eqnarray*}
where we used notations accepted above.
As a result of the validity of this equality we establish that the theorem is true.

If $b^0\in C_{\gamma}$, then the sequence $b(t)=(b_0,b_1(t),\ldots,b_{s}(t),\ldots)$
of the limit marginal observables determined by asymptotics (\ref{Iterd}) is generalized
global in time solution of corresponding initial-value problem of the dual Vlasov hierarchy:
\begin{eqnarray}\label{vdh}
   &&\hskip-9mm \frac{\partial}{\partial t}b_{s}(t)=
     \sum\limits_{j=1}^{s}\Lambda^{[1]}(j)\,b_{s}(t)
     +\sum_{j_1\neq j_{2}=1}^s\Lambda^{[2]}(j_1,j_{2})\,b_{s-1}(t,\textbf{u}_1,\ldots,\textbf{u}_{j_{2}-1},
       \textbf{u}_{j_{2}+1},\ldots,\textbf{u}_s),\quad s\geq1,
\end{eqnarray}
where in recurrence evolution equations (\ref{vdh}) the operators $\Lambda^{[1]}(j)$ and
$\Lambda^{[2]}(j_1,j_{2})$ are defined by formula (\ref{gen_obs_gener}) in case of $M=2$.

A similar approach to the description of kinetic evolution of quantum large particle systems
was considered in paper \cite{G11}.

\subsection{Relationships of marginal observables and marginal distribution functions}
\label{sec:34}
We consider initial states specified by a single-particle
marginal distribution function in the presence of correlations, namely
\begin{eqnarray}\label{lins}
   &&\hskip-8mm f^{(c)}\equiv(1,f_1^0(\textbf{u}_1),g_{2}(\textbf{u}_1,\textbf{u}_2)
        {\prod\limits}_{i=1}^{2}f_{1}^0(\textbf{u}_i),\ldots,
        g_{s}(\textbf{u}_1,\ldots,\textbf{u}_s){\prod\limits}_{i=1}^{s}f_{1}^0(\textbf{u}_i),\ldots),
\end{eqnarray}
where the bounded functions $g_{s}\equiv g_{s}(\textbf{u}_1,\ldots,\textbf{u}_s),\,s\geq2$, are
specified initial correlations\footnote{
We remark that correlations are usually described by means of the marginal correlation functions
defined by the cluster expansions of the marginal distribution functions ~\cite{GP},~\cite{G12}
\begin{eqnarray*}
   &&f_{s}^{0}(\textbf{u}_1,\ldots,\textbf{u}_s)=
     \sum\limits_{\mbox{\scriptsize$\begin{array}{c}\mathrm{P}:(\textbf{u}_1,\ldots,\textbf{u}_s)=\bigcup_{i}X_{i}\end{array}$}}
     \prod_{X_i\subset \mathrm{P}}g_{|X_i|}^{0}(X_i),\quad s\geq1,
\end{eqnarray*}
where ${\sum\limits}_{\mathrm{P}:(\textbf{u}_1,\ldots,\textbf{u}_s)=\bigcup_{i} X_{i}}$ is the sum over
all possible partitions $\mathrm{P}$ of the set $(\textbf{u}_1,\ldots,\textbf{u}_s)$ into $|\mathrm{P}|$
nonempty mutually disjoint subsets $X_i$. Hereupon solutions of these cluster expansions
\begin{eqnarray*}
   &&g_{s}^{0}(\textbf{u}_1,\ldots,\textbf{u}_s)=
      \sum\limits_{\mbox{\scriptsize $\begin{array}{c}\mathrm{P}:(\textbf{u}_1,\ldots,\textbf{u}_s)=\bigcup_{i}X_{i}\end{array}$}}
      (-1)^{|\mathrm{P}|-1}(|\mathrm{P}|-1)!\,\prod_{X_i\subset \mathrm{P}}f_{|X_i|}^{0}(X_i), \quad s\geq1,
\end{eqnarray*}
are interpreted as the functions that describe correlations of large particle systems.
If we consider initial states specified by the single-particle distribution
functions $f_1^{0}\equiv g_1^{0}$ in the presence of correlations, i.e.
initial states specified by the following sequence of marginal correlation functions
\begin{eqnarray*}
   &&\hskip-8mm g^{c}=\big(1,g_1^{0}(\textbf{u}_1),\tilde{g}_{2}(\textbf{u}_1,\textbf{u}_2)
        {\prod\limits}_{i=1}^{2}g_1^{0}(\textbf{u}_i),\ldots,
        \tilde{g}_{n}(\textbf{u}_1,\ldots,\textbf{u}_n){\prod\limits}_{i=1}^{n}g_1^{0}(\textbf{u}_i),\ldots\big),
\end{eqnarray*}
then the initial correlations of marginal distribution functions (\ref{lins}) and the marginal correlation
functions are expressed by the relations
\begin{eqnarray*}
   &&g_{s}(\textbf{u}_1,\ldots,\textbf{u}_s)=
          \sum\limits_{\mbox{\scriptsize $\begin{array}{c}\mathrm{P}:(\textbf{u}_1,\ldots,\textbf{u}_s)
          =\bigcup_{i}X_{i}\end{array}$}}\prod_{X_i\subset \mathrm{P}}\tilde{g}_{|X_i|},\quad s\geq2,
\end{eqnarray*}
and conversely,
\begin{eqnarray*}
   &&\tilde{g}_{s}(\textbf{u}_1,\ldots,\textbf{u}_s)=
      \sum\limits_{\mbox{\scriptsize $\begin{array}{c}\mathrm{P}:(\textbf{u}_1,\ldots,\textbf{u}_s)=\bigcup_{i}X_{i}\end{array}$}}
      (-1)^{|\mathrm{P}|-1}(|\mathrm{P}|-1)!\,\prod_{X_i\subset\mathrm{P}}g_{|X_i|}, \quad s\geq2.
\end{eqnarray*}
}.
Such states are intrinsic for the kinetic description of many-entity
systems in condensed states (for quantum large particle systems see papers \cite{GTsm},\cite{G14}).

If $b(t)\in C_{\gamma}$ and $f_1^0\in L^{1}(\mathcal{J}\times\mathcal{U})$,
then under the condition that $\|f_1^0\|_{L^{1}(\mathcal{J}\times\mathcal{U})}<\gamma$,
there exists the mean field scaling limit of the mean value functional of marginal
observables and it is determined by the following series expansion:
\begin{eqnarray*}
   &&\hskip-7mm\big(b(t),f^{(c)}\big)=\sum\limits_{s=0}^{\infty}\,\frac{1}{s!}\,
      \int_{(\mathcal{J}\times\mathcal{U})^s}d\textbf{u}_{1}\ldots d\textbf{u}_{s}
      \,b_{s}(t,\textbf{u}_1,\ldots,\textbf{u}_s)g_{s}(\textbf{u}_1,\ldots,\textbf{u}_s)
      \prod\limits_{i=1}^{s} f_1^0(\textbf{u}_i).
\end{eqnarray*}

Then for the mean value functionals of the limit initial additive-type marginal observables,
i.e. of the sequences $b^{(1)}(0)=(0,b_{1}^0(\textbf{u}_1),0,\ldots)$ ~\cite{BG10}, the
following representation is true:
\begin{eqnarray}\label{avmar-2}
   &&\hskip-9mm\big(b^{(1)}(t),f^{(c)}\big)=\sum\limits_{s=0}^{\infty}\,\frac{1}{s!}\,
       \int_{(\mathcal{J}\times\mathcal{U})^s}d\textbf{u}_{1}\ldots d\textbf{u}_{s}
       \,b_{s}^{(1)}(t,\textbf{u}_1,\ldots,\textbf{u}_s)g_{s}(\textbf{u}_1,\ldots,\textbf{u}_s)
       \prod\limits_{i=1}^{s} f_{1}^0(\textbf{u}_i)=\\
   &&=\int_{(\mathcal{J}\times\mathcal{U})}d\textbf{u}_{1}\,
       b_{1}^{0}(\textbf{u}_{1})f_{1}(t,\textbf{u}_{1}).\nonumber
\end{eqnarray}
In equality (\ref{avmar-2}) the function $b_{s}^{(1)}(t)$ is given by a special case of
expansion (\ref{Iterd}), namely
\begin{eqnarray*}\label{itvad}
   &&\hskip-9mm b_{s}^{(1)}(t,\textbf{u}_1,\ldots,\textbf{u}_s)=\\
   &&=\int_0^t dt_{1}\ldots\int_0^{t_{s-2}}dt_{s-1}
       \,e^{(t-t_{1})\sum\limits_{k_{1}=1}^{s}\Lambda^{[1]}(k_{1})}
       \sum\limits_{i_{1}\neq j_{1}=1}^{s}\Lambda^{[2]}(i_{1},j_{1})\,
       e^{(t_{1}-t_{2})\sum\limits_{l_{1}=1,\,\,l_{1}\neq j_{1}}^{s}\Lambda^{[1]}(l_{1})}\\
   &&\ldots\,e^{(t_{s-2}-t_{s-1})\sum\limits_{k_{s-1}=1,\,\,k_{s-1}\neq (j_{1},\ldots,j_{s-2})}^{s}
       \Lambda^{[1]}(k_{s-1})}\sum\limits^{s}_{\mbox{\scriptsize $\begin{array}{c}i_{s-1}\neq j_{s-1}=1,\\
       i_{s-1},j_{s-1}\neq (j_{1},\ldots,j_{s-2})\end{array}$}}\Lambda^{[2]}(i_{s-1},j_{s-1})\\
   &&\times e^{t_{s-1}\sum\limits_{l_{s-1}=1,\,\,l_{s-1}\neq (j_{1},\ldots,j_{s-1})}^{s}
       \Lambda^{[1]}(l_{s-1})}\,
       b_{1}^{0}((\textbf{u}_1,\ldots,\textbf{u}_s)\setminus(\textbf{u}_{j_{1}},\ldots,\textbf{u}_{j_{s-1}})),
       \quad s\geq1,
\end{eqnarray*}
and the limit single-particle distribution function $f_{1}(t)$ is represented by the series expansion
\begin{eqnarray}\label{viter}
   &&\hskip-9mm f_{1}(t,\textbf{u}_1)=\sum\limits_{n=0}^{\infty}\int_0^tdt_{1}\ldots\int_0^{t_{n-1}}dt_{n}
      \int_{(\mathcal{J}\times\mathcal{U})^n}d \textbf{u}_{2}\ldots d \textbf{u}_{n+1}\,
      e^{(t-t_{1})\Lambda^{\ast[1]}(1)}\Lambda^{\ast[2]}(1,2)\times\\
   &&\times\prod\limits_{j_1=1}^{2}e^{(t_{1}-t_{2})\Lambda^{\ast[1]}(j_1)}\ldots
      \prod\limits_{j_{n-1}=1}^{n}e^{(t_{n-1}-t_{n})\Lambda^{\ast[1]}(j_{n-1})}\times\nonumber\\
   &&\times\sum\limits_{i_{n}=1}^{n}\Lambda^{\ast[2]}(i_{n},n+1)
      \prod\limits_{j_n=1}^{n+1}e^{t_{n}\Lambda^{\ast[1]}(j_{n})}
      g_{1+n}(\textbf{u}_1,\ldots,\textbf{u}_{n+1})\prod\limits_{i=1}^{n+1}f_{1}^0(\textbf{u}_i),\nonumber
\end{eqnarray}
where the operators $\Lambda^{\ast[i]},\,i=1,2,$ are adjoint operators to the operators
$\Lambda^{[i]},\,i=1,2$ defined by formula (\ref{gen_obs_gener}), and on the space $L^{1}_{n}$
these operators are defined as follows:
\begin{eqnarray*}
  &&\hskip-8mm \Lambda^{\ast[1]}(i) f_n(\textbf{u}_1,\ldots,\textbf{u}_n)\doteq
     \int_{\mathcal{J}\times\mathcal{U}}
     A^{[1]}(\textbf{u}_{i};\textbf{v})a^{[1]}(\textbf{v})f_n(\textbf{u}_1,\ldots,
     \textbf{u}_{{i}-1},\textbf{v},\textbf{u}_{{i}+1},\ldots,\textbf{u}_n)d\textbf{v}-\\
  &&\hskip+9mm-a^{[1]}(\textbf{u}_{i})f_n(\textbf{u}_1,\ldots,\textbf{u}_n),
\end{eqnarray*}
\begin{eqnarray*}
  &&\hskip-8mm \Lambda^{\ast[2]}(i,j)f_n(\textbf{u}_1,\ldots,\textbf{u}_n)\doteq
    \int_{\mathcal{J}\times\mathcal{U}} A^{[2]}(\textbf{u}_{i};\textbf{v},\textbf{u}_{j})
     a^{[2]}(\textbf{v},\textbf{u}_{j})f_n(\textbf{u}_1,\ldots,\textbf{u}_{{i}-1},\textbf{v},
     \textbf{u}_{{i}+1},\ldots,\textbf{u}_n)d\textbf{v}-\\
  &&\hskip+9mm-a^{[2]}(\textbf{u}_{i},\textbf{u}_{j})f_n(\textbf{u}_1,\ldots,\textbf{u}_n)\nonumber,
\end{eqnarray*}
where the functions $A^{[m]},a^{[m]},\,m=1,2$, are defined above in formula (\ref{gen_obs_gener}).

For initial data $f_{1}^0\in L^{1}(\mathcal{J}\times\mathcal{U})$ limit marginal distribution
function (\ref{viter}) is a strong solution of the Cauchy problem of the Vlasov-type kinetic equation
with initial correlations:
\begin{eqnarray}
  \label{Vlasov1}
    &&\hskip-5mm\frac{\partial}{\partial t}f_{1}(t,\textbf{u}_1)= \Lambda^{\ast[1]}(1)f_{1}(t,\textbf{u}_1)+\\
    &&\hskip+5mm+\int_{\mathcal{J}\times\mathcal{U}}d\textbf{u}_{2}
       \Lambda^{\ast[2]}(1,2)\prod_{i_1=1}^{2}e^{t\Lambda^{\ast[1]}(i_1)}g_{2}(\textbf{u}_1,\textbf{u}_2)
     \prod_{i_2=1}^{2}e^{t\Lambda^{\ast[1]}(i_2)}
     f_{1}(t,\textbf{u}_1)f_{1}(t,\textbf{u}_2),\nonumber\\ \nonumber\\
  \label{Vlasov2}
    &&\hskip-5mmf_{1}(t,\textbf{u}_1)|_{t=0}=f_1^0(\textbf{u}_1), \nonumber
\end{eqnarray}
where the function $g_{2}(\textbf{u}_1,\textbf{u}_2)$ is initial
correlation function ~\cite{GP} specified initial data (\ref{lins}).

For the mean value functionals of the limit initial $k$-ary marginal observables, i.e. of
the sequences $b^{(k)}(0)=(0,\ldots,0,b_{k}^0(\textbf{u}_1,\ldots,\textbf{u}_k),0,\ldots),\,k\geq2$,
the following equality is true:
\begin{eqnarray}\label{dchaos}
    &&\hskip-8mm\big(b^{(k)}(t),f^{(c)}\big)=\sum\limits_{s=0}^{\infty}\,\frac{1}{s!}\,
       \int_{(\mathcal{J}\times\mathcal{U})^s}d\textbf{u}_{1}\ldots d\textbf{u}_{s}
       \,b_{s}^{(k)}(t,\textbf{u}_1,\ldots,\textbf{u}_s) g_{s}(\textbf{u}_1,\ldots,\textbf{u}_s)
       \prod\limits_{i=1}^{s} f_1^0(\textbf{u}_i)=\\
    &&=\frac{1}{k!}\int_{(\mathcal{J}\times\mathcal{U})^k}d\textbf{u}_{1}\ldots d\textbf{u}_{k}
       \,b_{k}^0(\textbf{u}_1,\ldots,\textbf{u}_k)\prod_{i_1=1}^{k}e^{t\Lambda^{\ast[1]}(i_1)}
       g_{k}(\textbf{u}_1,\ldots,\textbf{u}_k)\prod_{i_2=1}^{k}e^{t\Lambda^{\ast[1]}(i_2)}
       \prod\limits_{i=1}^{k} f_{1}(t,\textbf{u}_i),\nonumber
\end{eqnarray}
where the limit single-particle marginal distribution function $f_{1}(t,\textbf{u}_i)$ is determined
by series expansion (\ref{viter}) and the function $g_{k}(\textbf{u}_1,\ldots,\textbf{u}_k)$
is initial correlation function specified initial states (\ref{lins}).
Hence in case of the $k$-ary marginal observables the evolution governed by the dual Vlasov
hierarchy (\ref{vdh}) is equivalent to a property of the propagation of initial correlations
(\ref{dchaos}) for the $k$-particle marginal distribution functions or in other words the mean
field scaling dynamics does not create correlations.

Thus, an equivalent approach to the description of the kinetic evolution of large number
of interacting constituents in terms of the Vlasov-type kinetic equation with correlations
(\ref{Vlasov1}) is given by the dual Vlasov hierarchy (\ref{vdh}) for the additive-type
marginal observables.

\section{Conclusion}
\label{sec:4}
The properties of semigroups of the theory of hierarchies of evolution equations of large 
particle systems in suitable functional spaces were considered. It was established that 
the generating operators of the expansions for one-parameter families of operators of these 
hierarchies are the corresponding order cumulants (semi-invariants) of semigroups of the 
Liouville equations for states or observables.

In this paper we also considered the possible application of obtained results to the description
of kinetic evolution of large number of interacting constituents of soft active matter within
the framework of marginal observables governed by the dual BBGKY hierarchy. Such representation
of the kinetic evolution seems, in fact, the direct mathematically fully consistent formulation,
modeling collective behavior of biological systems, since the notion of state is more subtle and
implicit characteristic of living entities.

One of the advantages of the developed approach is an opportunity to derive the kinetic equations
with initial correlations that may characterize the condensed states of large particle systems.

We note that properties of the corresponding semigroups in case of quantum many-particle systems
were considered in review \cite{G12} and the general approaches to the description of the evolution
of states within the framework of correlation operators and marginal correlation operators were
developed in paper ~\cite{GP}.


\begin{thebibliography}{99.}%

\bibitem{BanArl}
               Banasiak J., Arlotti L.:
               Perturbations of Positive Semigroups with Applications. Springer-Verlag, London (2006)

\bibitem{BM}
               Belleni-Morante A., McBride A.C.:
               Applied Nonlinear Semigroups: An Introduction. John Wiley and Sons, Inc., Chichester (1998)

\bibitem{CGP97}
               Cercignani C., Gerasimenko V.I., Petrina D.Ya.:
               Many-Particle Dynamics and Kinetic Equations. Kluwer Acad. Publ., Dordrecht (1997)

\bibitem{SR12}
               Gallagher I., Saint-Raymond L., Texier B. From Newton to Boltzmann: Hard Spheres and Short-range Potentials.
               EMS Publ. House: Z\"{u}rich Lectures in Advanced Mathematics (2014)

% Journal article
\bibitem{G}
               Gerasimenko V.I.: On the approaches to the derivation of the Boltzmann equation with hard
               sphere collisions.
               Proc. Inst. Math. NASU. \textbf{10}(2), 71--95 (2013)

\bibitem{GF12}
               Gerasimenko V.I., Fedchun Yu.Yu.:
               Nonperturbative solution expansions of hierarchies of evolution equations in functional derivatives.
               Proc. Inst. Math. NASU. \textbf{9}(2), 347--375 (2012)

\bibitem{L11}
               Lachowicz M.: Individually-based Markov processes modeling nonlinear systems in mathematical biology.
               Nonlinear Analysis: Real World Applications. \textbf{12}, 2396--2408 (2011)

\bibitem{MJRLPRAS}
               Marchetti M.C., Joanny J.F., Ramaswamy S., Liverpool T.B., Prost J., Rao M., Simha R.A.:
               Hydrodynamics of soft active matter. Rev. Mod. Phys. \textbf{85}, 1143--1194, (2013)

\bibitem{ALBA}
               Bellouquid A., Delitala M.:
               Mathematical Modeling of Complex Biological Systems: A Kinetic Theory Approach.
               Birkh\"{a}user, Boston (2006)

\bibitem{GF13}
               Gerasimenko V.I., Fedchun Yu.Yu.: On kinetic models for the evolution of many-entity systems
               in mathematical biology.
               J. Coupled Syst. Multiscale Dyn. \textbf{1}(2), 273--279 (2013)

\bibitem{BG10}
               Borgioli G., Gerasimenko V.I.: Initial-value problem of the quantum dual BBGKY hierarchy.
               Nuovo Cimento Soc. Ital. Fis. C. \textbf{33}(1),  71--78 (2010)

\bibitem{BG}
               Borovchenkova M.S., Gerasimenko V.I.: On the non-Markovian Enskog equation for granular gases.
               J. Phys. A: Math. Theor. \textbf{47}(3), 035001 (2014)

\bibitem{G12}
               Gerasimenko V.I.: Hierarchies of quantum evolution equations and dynamics of many-particle correlations.
               In: Statistical Mechanics and Random Walks: Principles, Processes and Applications.
               Nova Science Publ., Inc., N.Y., 233--288 (2012)

\bibitem{GF14}
               Gerasimenko V.I., Fedchun Yu.Yu.: Kinetic equations of soft active matter.
               Reports NAS of Ukraine. \textbf{5}, 11--18 (2014)

\bibitem{BL14}
               Banasiak J., Lachowicz M.: Methods of Small Parameter in Mathematical Biology.
               Boston, Birkh\"{a}user (2014)

\bibitem{G11}
               Gerasimenko V.I.: Heisenberg picture of quantum kinetic evolution in mean-field limit.
               Kinet. Relat. Models. \textbf{4}(1), 385--399 (2011)

\bibitem{GP}
               Gerasimenko V.I., Polishchuk D.O.: A nonperturbative solution of the nonlinear BBGKY hierarchy
               for marginal correlation operators.
               Math. Methods Appl. Sci. \textbf{36}(17), 2311--2328 (2013)

\bibitem{GTsm}
               Gerasimenko V.I., Tsvir Zh.A.: On quantum kinetic equations of many-particle systems
               in condensed states.
               Physica A: Stat. Mech. Appl. \textbf{391}(24), 6362--366 (2012)

\bibitem{G14}
               Gerasimenko V.I.: Kinetic evolution of quantum large particle systems with initial
               correlations.
               Preprint arXiv:1407.1633 [math.AP], (2014)

\end{thebibliography}
\end{document}